\documentclass[sigconf]{acmart}

\AtBeginDocument{%
  }

\setcopyright{acmcopyright}
\copyrightyear{2024}
\acmYear{2024}
\acmDOI{XXXXXXX.XXXXXXX}

\usepackage{enumitem}
\usepackage{multirow}
\usepackage{bbding}
\usepackage{bm}
\usepackage{xcolor}

\setitemize[1]{itemsep=0pt,partopsep=0pt,parsep=\parskip,topsep=0pt}

\begin{document}

\title{Behavior-Contextualized Item Preference Modeling for Multi-Behavior Recommendation}

\author{Mingshi Yan}
\affiliation{%
  \institution{College of Intelligence and Computing, Tianjin University}
  \country{China}
}
\email{neo.ms.yan@gmail.com}

\author{Fan Liu}
\affiliation{%
  \institution{School of Computing, National University of Singapore}
  \country{Singapore}
}
\email{liufancs@gmail.com}

\author{Jing Sun}
\affiliation{%
  \institution{School of Information and Communication Engineering, Dalian Minzu University}
  \country{China}
}
\email{jingsun@dlnu.edu.cn}

\author{Fuming Sun}
\affiliation{%
  \institution{School of Information and Communication Engineering, Dalian Minzu University}
  \country{China}
}
\email{sunfuming@dlnu.edu.cn}

\author{Zhiyong Cheng}
\affiliation{%
  \institution{School of Computer Science and Information Engineering, Hefei University of Technology}
  \country{China}
}
\email{jason.zy.cheng@gmail.com}

\author{Yahong Han}
\authornote{Y. Han is the corresponding author.}
\affiliation{%
  \institution{College of Intelligence and Computing, Tianjin University}
  \country{China}
}
\email{yahong@tju.edu.cn}

\renewcommand{\shortauthors}{Mingshi Yan et al.}

\begin{abstract}

   In recommender systems, multi-behavior methods have demonstrated their effectiveness in mitigating issues like data sparsity, a common challenge in traditional single-behavior recommendation approaches. These methods typically infer user preferences from various auxiliary behaviors and apply them to the target behavior for recommendations. However, this direct transfer can introduce noise to the target behavior in recommendation, due to variations in user attention across different behaviors. To address this issue, this paper introduces a novel approach, Behavior-Contextualized Item Preference Modeling (BCIPM), for multi-behavior recommendation. Our proposed Behavior-Contextualized Item Preference Network discerns and learns users' specific item preferences within each behavior. It then considers only those preferences relevant to the target behavior for final recommendations, significantly reducing noise from auxiliary behaviors.  These auxiliary behaviors are utilized solely for training the network parameters, thereby refining the learning process without compromising the accuracy of the target behavior recommendations. To further enhance the effectiveness of BCIPM, we adopt a strategy of pre-training the initial embeddings. This step is crucial for enriching the item-aware preferences, particularly in scenarios where data related to the target behavior is sparse. Comprehensive experiments conducted on four real-world datasets demonstrate BCIPM's superior performance compared to several leading state-of-the-art models, validating the robustness and efficiency of our proposed approach.
   

\end{abstract}



\maketitle

\section{Introduction}

With the rapid advancement of the Internet, individuals are inundated with an ever-increasing volume of information and choices. Recommender systems~\cite{NCF, wide&deep} provide an effective way to filter and curate information, thereby saving users time and effort. 
Collaborative Filtering (CF)~\cite{SuK09, DongZSLC18, LightGCN, Liu2021IMP_GCN, wei2023lightgt} plays a pivotal role in recommender systems. CF operates by relying solely on user-item interaction data to learn user and item embeddings, which respectively represent user preferences and item features. It then employs a similarity function to estimate the compatibility between users and items based on their embeddings to provide recommendations. Traditional CF methods, typically focusing on a single behavior such as \textit{purchase}, often encounter challenges with data sparsity, leading to dramatic performance degradation.   

In real scenarios,  user interactions extend beyond \textit{purchase} to include behaviors like \textit{click} and \textit{cart}, which also provide rich insights into user preferences. To overcome the challenges posed by data sparsity, researchers have introduced  Multi-behavior Recommendation (MBR) methods~\cite{MBGCN, NMTR, EHCF}, which exploit data from multiple user behaviors to learn user preferences.
In these models, while the most desired behavior, typically \textit{purchase}, is considered the target behavior, other user interactions such as \textit{click} and \textit{cart} are utilized as auxiliary behaviors. This approach helps in creating a more robust and accurate recommendation system by utilizing a broader spectrum of user data.

Recent advancements in MBR methods have marked significant progress, as evidenced by various studies \cite{NMTR, HMG-CR, KHGT, PKEF}. Initially, early matrix factorization-based approaches \cite{CMF, BF} utilized shared embeddings to extend matrix factorization across multiple behaviors. Subsequently, with the evolution of Deep Neural Networks (DNN) and Graph Convolutional Networks (GCN), a new wave of methods emerged, combining CF with these advanced technologies.  One approach treats multiple behaviors as distinct spaces~\cite{ARGO, EHCF}, employing DNN to learn user preferences for each behavior individually. Techniques such as projection are then used to transfer user preferences from auxiliary behaviors to the target behavior space for fusion. Another approach involves utilizing GCN to perform graph convolutions on each behavior separately~\cite{MBGCN, SMBREC} and then fusing the learned user representations for target behavior recommendation. Despite the technical difference, these methods share a common goal: incorporating user preferences from auxiliary behaviors into the target behavior to enhance user representation.

\begin{figure}[]
    \centering
    \includegraphics[width=1.\linewidth]{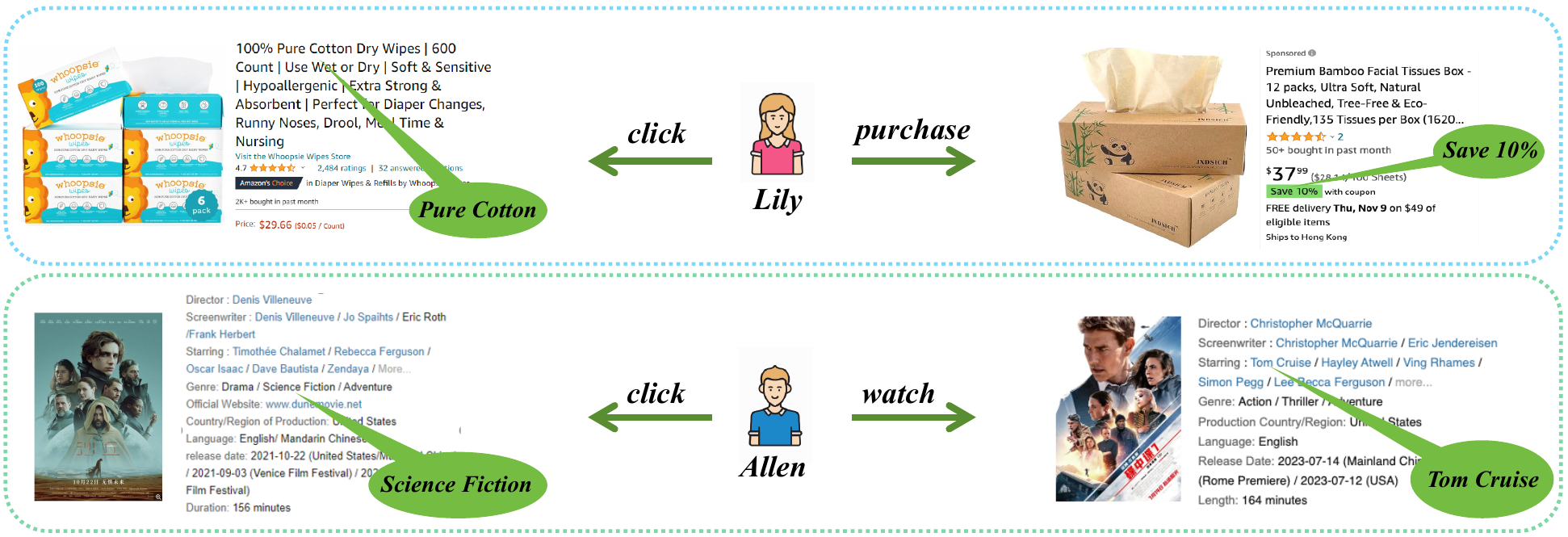}
    \caption{Examples of user behavior influenced by their decision factors.}
    \label{fig:fig1}
\end{figure}

While current MBR methods have shown effectiveness, they face a significant challenge: the direct information transfer from the auxiliary behaviors to the target behaviors may introduce noise due to the variation of user focus across different behaviors. This issue can be vividly illustrated through real-world scenarios.  As depicted in Figure~\ref{fig:fig1}, consider the case of Lily, who, when purchasing tissues, is initially drawn to the "\textit{pure cotton}" for tissues. However, she ultimately chooses a different brand,  influenced by a 10\% cost-saving. Similarly, Allen, on a movie website, initially clicks on "\textit{Dune}" due to his interest in the "\textit{science fiction}" genre, but eventually watches "\textit{Mission Impossible}" for its star "\textit{Tom Cruise}". These examples highlight that user interactions across various behaviors are influenced by different aspects of item features. When preferences learned from auxiliary behaviors are merged into the target behavior, they enrich the user's profile but also introduce deviations from their actual preferences in the target behavior.  This is attributed to differences in user focus between behaviors. As a result, the partial preference information gleaned from other behaviors can inadvertently introduce noise when applied to the context of the target behavior.
It's important to recognize that this variation in user focus is prevalent not only across different behaviors but also within the same behavior, influenced by the specific characteristics of each item. We term these influencing factors as \textbf{item-aware preference}. Acknowledging these item-aware preferences allows for a deeper understanding of users' specific points of interest in their interactions. However, current CF-based recommendation methods tend to overlook these nuanced preferences. Our approach aims to address this gap by considering item-aware preferences, thereby providing a more accurate representation of user interests and enhancing recommendation effectiveness.

In addressing the challenges identified within MBR systems, we propose a novel approach, Behavior-Contextualized Item Preference Modeling (BCIPM), for Multi-Behavior Recommendation. The core of this approach is the Behavior-Contextualized Item Preference Network (BIPN), which extracts item-specific preferences from user-item interactions of various behaviors. BIPN features a three-layer architecture that models the interactions between users and items within specific behaviors. A key feature of BIPN is its strategic use of data from auxiliary behaviors, which is exclusively employed for training the network's parameters. This ensures that auxiliary behaviors do not exert a direct influence on the final recommendation. Our model generates recommendations by aggregating only item-aware preferences relevant to users' target behavior, effectively bypassing the noise introduced by auxiliary behaviors. As random embeddings often lack informative value, we pre-train user and item embeddings via the GCN-based method. This pre-training captures a wider range of collaborative information by using multi-behavior interaction data without distinguishing between behavior types, enriching the initial embeddings with more relevant data. Moreover, to tackle the data sparse issue in the target behavior, which may lead to insufficient item-aware preferences, we introduce a GCN enhancement module. This module is designed to reinforce user preferences specifically within the target behavior, ensuring a more robust and accurate recommendation. We conduct extensive experiments on four different real-world datasets, and our proposed BCIPM significantly surpasses existing methods. The experimental results also demonstrate the effectiveness of BCIPM in reducing the noise from auxiliary behaviors in multi-behavior recommendation systems.

In summary, the main contribution of this work is threefold:
\begin{itemize}[leftmargin=*]
    
    \item We introduce the Behavior-Contextualized Item Preference Network, an innovative network designed to distill item-aware preferences from user-item interactions. This network is finely tuned to model the nuanced interactions between individual users and items within specific behaviors.
    
    \item We employ GCN to pre-train the initial embeddings with multi-behavior interaction data. This technique allows our model to harness a richer set of collaborative information from these interactions, thereby significantly boosting the learning and accuracy of item-aware preferences.

    \item Our approach has been rigorously tested through extensive experiments conducted on four real-world datasets. The results consistently demonstrate the superior effectiveness and generality of our model. Compared to state-of-the-art methods, BCIPM shows notable performance improvements, underscoring its practical applicability and robustness in diverse scenarios. Our code is available on GitHub~\footnote{https://github.com/MingshiYan/BIPN} for research purpose.
     
  \end{itemize}

\section{Related Work} \label{Related Work}

Recently, MBR~\cite{BPRH, EHCF, HMG-CR, CRGCN} methods have gained significant attention for their effectiveness in mitigating data sparsity issues. Based on technological evolution, MBR methods can be broadly classified into three categories: traditional machine learning-based, DNN-based, and GCN-based approaches.

Traditional machine learning-based methods, constituting the initial phase of MBR approaches, primarily focus on enhancing user and item representation learning through data derived from auxiliary behaviors. One such method utilizes matrix factorization~\cite{CMF, BF, BPRH}, designed to refine representation learning by factoring matrices of auxiliary behaviors using shared embeddings. 
Another strategy involves sampling interaction data from auxiliary behaviors~\cite{LoniPLH16, DingY0QLCJY18, GuoQTLMW17} to enrich the primary behavior dataset, thus facilitating a better understanding of user preferences. 

DNN-based methods~\cite{RCF, MATN, ARGO} integrate collaborative filtering with deep neural networks. These methods exploit the robust ability of neural networks to process complex, nonlinear relationships, thereby uncovering deeper latent associations between users and items. 
For instance, Xin et al.~\cite{RCF} combined neural networks and attention mechanisms to model item-based collaborative filtering networks and relation-based collaborative filtering networks separately. 
Moreover, within the DNN framework, these methods have been combined with other recent technological advancements~\cite{NMTR, EHCF}. 

GCN-based methods~\cite{GNMR, MBGCN, SMBREC, CKML} leverage GCNs to model collaborative information between users and items. The strength of GCNs in representing graph-structured data has proven advantageous in this context. A primary method, proposed by Jin et al.~\cite{MBGCN}, involves applying GCNs to learn user preferences within each behavioral graph, followed by an aggregation process. Moreover, GCN-based approaches have incorporated advanced techniques~\cite{PKEF, MBSSL, GHCF, HMG-CR}. For example, Luo et al.~\cite{GHCF} implemented a non-sampling strategy and developed a relation-aware GCN propagation layer through multi-task learning, embedding both node representations and relationships within the graph.

While these MBR methods vary in their technical approaches, they commonly rely on collaborative information from auxiliary behaviors to enhance user representation learning. Traditional machine learning-based methods derive this information directly from user-item interactions within auxiliary behaviors. In contrast, DNN and GCN-based methods model the overarching auxiliary behavior. However, these approaches tend to overlook the potential noise introduced by auxiliary behaviors. Our proposed method contrasts with these by focusing on enhancing the accuracy of user preference learning through item-aware preference modeling within specific user behaviors. In our approach, auxiliary behavior data is utilized solely for embedding pre-training and network parameter optimization, effectively avoiding the introduction of noise.

\section{methodology} \label{methodology}

\subsection{Preliminaries}


In this section, we will provide a detailed description of the BCIPM model. Before delving into that, we first introduce the problem under investigation.
Giving $\mathcal{U}=\{u_{1}, \cdots, u_{m}, \cdots, u_{M}\}$ and $\mathcal{I}=\{i_{1}, \cdots, i_{n}, \cdots, i_{N}\}$ as the set of users and items, where $M$ and $N$ denote the total number of users and items, with $u_{m}$ and $i_n$ representing the $m$-$th$ user and $n$-$th$ item, respectively. We introduce the matrices $\boldsymbol{P} \in \mathbb{R}^{M \times d}$ and $\boldsymbol{Q} \in \mathbb{R}^{N \times d}$, serving as the embedding matrices for users and items, where $d$ is the embedding dimension. The user and item embeddings, $\boldsymbol{e}_u^0$ and $\boldsymbol{e}_i^0$, are retrieved from $\boldsymbol{P}$ and $\boldsymbol{Q}$ based on identifiers from the sets $\mathcal{U}$ and $\mathcal{I}$, respectively, and serve as the initial embeddings for users and items. In addition, we define $\mathcal{B}=\{b_{1}, \cdots, b_{k}, \cdots, b_{K}\}$ as the set of behaviors, and $b_{K}$ is the target behavior. For the initialization of behaviors, we employ one-hot embedding $\boldsymbol{e}_b$ to indicate a specific behavior $b$.

Our objective is to optimize these user, item, and behavior embeddings to obtain user and item representation from multi-behavior interaction data. Subsequently, the model aims to output a list of recommended items for a target user, focusing specifically on the target behavior.


\subsection{Behavior-Contextualized Item Preference Modeling}
The motivation for this work stems from the observation that user interactions are often significantly influenced by item-specific characteristics. This insight leads us to conclude that aggregating these item-aware factors across different items in a specific behavior can provide an accurate representation of a user's preferences for that behavior. Inspired by this, we design a network that focuses on learning item-aware preference embeddings within a specific behavior. To enhance the representation of these item-aware preference embeddings, we begin with pre-training on the initial embeddings. This pre-training leverages data from multiple behaviors, allowing us to utilize a broader and more nuanced spectrum of user-item interaction data. This strategy is particularly crucial in addressing the data sparsity issues~\cite{liu2023semantic} in recommender systems. Moreover, we recognize that pre-training alone might not suffice in fully capturing the item-aware preference embeddings for the target behavior, which could result in sub-optimal recommendation performance. To mitigate this potential shortfall, we further refine and enhance the user preference embeddings specifically for the target behavior. 

\begin{figure}[t]
    \centering
    \includegraphics[width=\linewidth]{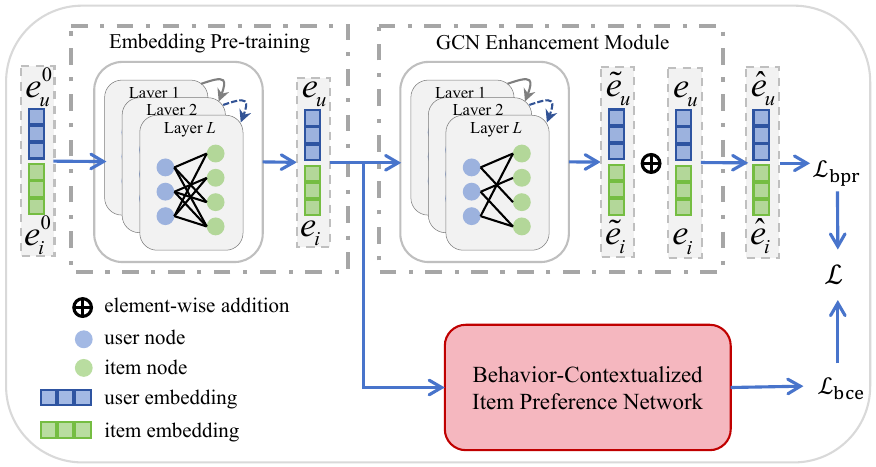}
    \caption{Overview of our proposed BCIPM.}
    \label{fig:global}
\end{figure}

As depicted in Figure~\ref{fig:global}, our model consists of three components: (1) \textbf{Embedding Pre-training}: This component is designed to pre-train the user and item embeddings based on the randomly initialized embeddings and collaborative information. (2) \textbf{Behavior-Contextualized Item Preference Network}: This is the core component of our model. It focuses on learning behavior-contextualized item preferences which capture the nuanced interplay between users' interactions and item characteristics within specific behaviors. (3) \textbf{GCN Enhancement Module}: This module plays a crucial role in enhancing user preferences, particularly targeting the potential issue caused by data sparsity in the target behavior. In the next, we will provide a detailed introduction to these three components.

\subsubsection{\textbf{Embedding Pre-training}} \label{Embedding Pre-training}

Prior to capturing item-aware preference, it is essential to acknowledge that user and item embeddings are typically initialized randomly using specific strategies (e.g., normal distribution or uniform distribution) and inherently lack valuable information. However, imbuing the embeddings with collaborative information is a prerequisite for learning the item-aware preference. Consequently, it becomes necessary to pre-train the embeddings before inputting them into the behavior-contextualized item preference network. 

Given its notable effectiveness in representation learning, we employ the GCN technique for embedding pre-training. In this paper, we do not distinguish between different types of behaviors, but instead apply graph convolution to a single unified graph created from interaction data encompassing all behaviors. It is justified by the idea that ignoring behavior types allows for learning fine-grained representations within a broader graph. Additionally, these representations are not influenced by behavioral factors (See Section~\ref{Strategy Analysis} for detailed experimental analysis).

Owing to the effectiveness of LightGCN~\cite{LightGCN}, it is utilized to perform graph convolution on the constructed unified graph. Specifically, we construct an interaction matrix $\boldsymbol{R} \in \mathbb{R}^{M \times N}$ with user-item interactions from all behaviors, where $M (N)$ represents the total number of users (items). When observing a user-item interaction in any behavior, the corresponding element $r_{ui}$ in the matrix $\boldsymbol{R}$ is set to 1, otherwise 0.
Here, we briefly recap the application of LightGCN on the interaction matrix $\boldsymbol{R}$,
\begin{equation}
  \label{eq:ajm}
   \boldsymbol{A} = \begin{bmatrix}
        \boldsymbol{0} & \boldsymbol{R} \\
        \boldsymbol{R}^{\top} & \boldsymbol{0}
    \end{bmatrix} ,
\end{equation}
where $\boldsymbol{A}$ is the adjacency matrix of the graph. Next, the information from high-order neighborhoods is aggregated through the graph convolution operation,
\begin{equation}
  \label{eq:gcn}
   \boldsymbol{E}^{(l+1)} = (\boldsymbol{D}^{-\frac{1}{2}} \boldsymbol{A} \boldsymbol{D}^{-\frac{1}{2}}) \boldsymbol{E}^{(l)},
\end{equation}
where $\boldsymbol{E}^{(l)}$ represents the $l$-$th$ layer node embedding, when $l=0$, the $\boldsymbol{E}^{(0)}$ is the initialization of the user and item embeddings, and $\boldsymbol{D}$ denotes the diagonal identity matrix. Finally, we aggregate the layer node embeddings,
\begin{equation}
  \label{eq:sum}
   \boldsymbol{E} = \sum_{l=0}^{L}{\alpha_l \boldsymbol{E}^{(l)}},
\end{equation}
where $\alpha$ is the weight coefficient, and following LightGCN, we set it as $1/(l+1)$. $\boldsymbol{E}$ represents the embeddings of users and items, serving as inputs for subsequent modules, with its embedding form denoted as $\boldsymbol{e}_u$ for user $u$ and $\boldsymbol{e}_i$ for item $i$.



\subsubsection{\textbf{Behavior-Contextualized Item Preference Network}} \label{Behavior-Contextualized Item Preference network}

\begin{figure}[]
    \centering
    \includegraphics[width=\linewidth]{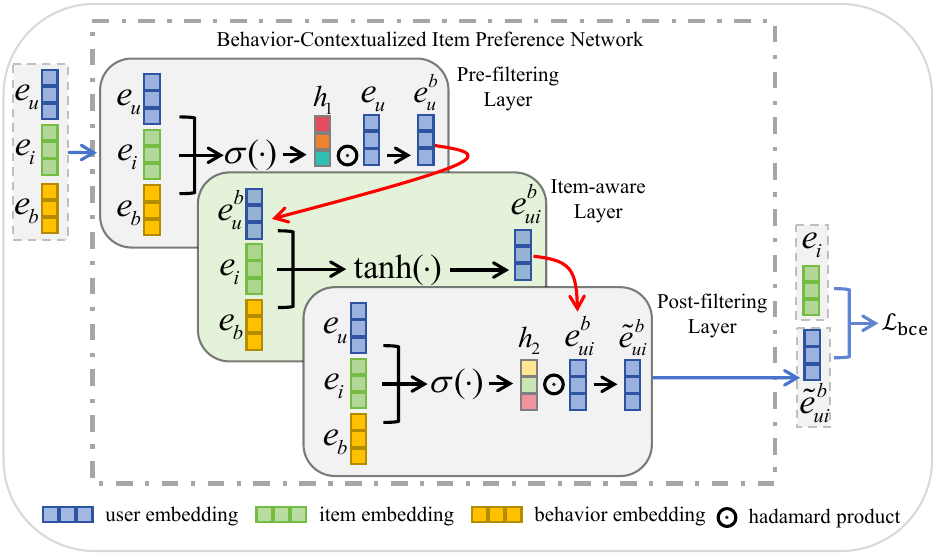}
    \caption{Structure of the Behavior-Contextualized Item Preference Network.}
    \label{fig:BIPM_net}
\end{figure}

The core of the BIPN, as shown in Figure~\ref{fig:BIPM_net}, is dedicated to unveiling users' item-aware preference in a specific behavior. This network is structured with a \textit{\textbf{Pre-filtering Layer}}, an \textit{\textbf{Item-aware Layer}}, and a \textit{\textbf{Post-filtering Layer}}. The network takes a triplet <$\boldsymbol{e}_u, \boldsymbol{e}_i, \boldsymbol{e}_b$> as input, where $\boldsymbol{e}_u$ and $\boldsymbol{e}_i$ denote the pre-trained embeddings of the user $u$ and item $i$, respectively, and $\boldsymbol{e}_b$ represents the embedding of the behavior $b$.

\textbf{Pre-filtering Layer.} This layer primarily functions to filter the input user embeddings. As discussed in the preceding section, the pre-trained embeddings are not influenced by behavioral factors. Therefore, it becomes imperative to filter user embeddings, retaining only relevant information aligned with the current behavior before learning item-aware preference in a specific behavior. To achieve this, triplet <$\boldsymbol{e}_u, \boldsymbol{e}_i, \boldsymbol{e}_b$> are employed to learn a weight vector $\boldsymbol{h}_1 \in \mathbb{R}^{d}$. This vector is used to perform element-wise multiplication on user embeddings, facilitating element-level information filtering~\cite{MEGAN},
\begin{equation}
  \label{w1}
    \boldsymbol{h}_1 = \sigma (\boldsymbol{W}_1 (\boldsymbol{e}_u || \boldsymbol{e}_i || \boldsymbol{e}_b) + \boldsymbol{b}_1) ,
\end{equation}
\begin{equation}
  \label{e_filter}
    \boldsymbol{e}_u^{b} = \boldsymbol{h}_1 \odot \boldsymbol{e}_u ,
\end{equation}
where $\boldsymbol{e}_u^{b}$ represents the preference information related to item $i$ that user $u$ may possess within behavior $b$, $\sigma$ is the sigmoid function, $\boldsymbol{W}_1 \in \mathbb{R}^{d \times (2d + l)}$ and $\boldsymbol{b}_1 \in \mathbb{R}^{d}$ are trainable parameters, $d$ and $l$ denote the embedding dimensions of the user (item) and behavior, respectively. $||$ means the concatenation operation, $\odot$ represents the Hadamard product.

\textbf{Item-aware Layer.} In this layer, we aim to capture the user's specific preference when interacting with a particular item in a behavior-contextualized setting. It can be formulated as:
\begin{equation}
  \label{eq:4}
    \boldsymbol{e}_{ui}^{b} = tanh (\boldsymbol{W}_2 (\boldsymbol{e}_u^{b} || \boldsymbol{e}_i || \boldsymbol{e}_b) + \boldsymbol{b}_2) ,
\end{equation}
where $\boldsymbol{e}_{ui}^{b}$ represents the item-aware preference of user $u$ for item $i$ in behavior $b$, $\boldsymbol{W}_2 \in \mathbb{R}^{d \times (2d + l)}$ and $\boldsymbol{b}_2 \in \mathbb{R}^{d}$ are trainable parameters.

\textbf{Post-filtering Layer.} In the final layer, we apply additional filtering to the learned preferences, similar to the pre-filter layer. This layer is designed to capture better item-aware behavior-specific preference, and it is formulated as:
\begin{equation}
  \label{w1}
    \boldsymbol{h}_2 = \sigma (\boldsymbol{W}_3 (\boldsymbol{e}_u || \boldsymbol{e}_i || \boldsymbol{e}_b) + \boldsymbol{b}_3) ,
\end{equation}
\begin{equation}
  \label{e_filter}
    \boldsymbol{\tilde{e}}_{ui}^{b} = \boldsymbol{h}_2 \odot \boldsymbol{e}_{ui}^{b},
\end{equation}
where $\boldsymbol{\tilde{e}}_{ui}^{b}$ represents the behavior-contextualized item preference embedding of user $u$ for item $i$ in behavior $b$.

Intuitively, the higher the correlation between item-aware preferences with the item, the more accurate the learning of item-aware preferences will be. Here, we adopt cross-entropy loss to optimize the network,
\begin{equation}
  \label{eq:bce}
   \mathcal{L}_{bce} = -\sum \ \ y \cdot log \sigma(y') + (1-y) \cdot log(1 - \sigma(y'),
\end{equation}
where $y$ is the ground truth, denoting 1 when the $u$-$i$ interaction is observed in behavior $b_k$ and 0 otherwise, $y' = (\boldsymbol{\tilde{e}}_{ui}^{b_k})^{\top} \boldsymbol{e}_i$, and $b_k \in \mathcal{B}$. 

\textbf{User's target behavior preference learning.} After training the BIPN with the interaction data of all behaviors, we then use the trained network to learn user preferences in the target behavior. Specifically, for a user $u$, we aggregate the item-aware preferences learned from all interacted items in the target behavior (denoted as $\mathcal{I}_{b_K}$) to represent the user's specific preference $\tilde{\boldsymbol{e}}_{u}^{agg}$ in the target behavior,
\begin{equation}
  \label{eq:agg}
   \tilde{\boldsymbol{e}}_{u}^{agg} = \sum_{i \in \mathcal{I}_{b_K}} \boldsymbol{\tilde{e}}_{ui}^{b_K},
\end{equation}
where $b_K$ refers to the target behavior.

\subsubsection{\textbf{GCN Enhancement Module}}
In the above sections, we learn user preference in the target behavior using the BIPN with the pre-trained embeddings, which are optimized using data of all behaviors. Note that BIPN is a preference-filtering network to identify item-aware preference, it is not a collaborative filtering model for a specific behavior. 
When the user's interaction records are insufficient in the target behavior, the aggregated user preference by Eq.~\ref{eq:agg} may fail to accurately reflect the user preferences.
To address this issue, we leverage GCN's ability to explore higher-order neighbor information to enhance user preferences. For simplicity, we directly adopt LightGCN. As described in Section~\ref{Embedding Pre-training}, with $\boldsymbol{E}$, which is the pre-trained embeddings of users and items, as the initialization, to perform graph convolution operation on the target behavior, resulting in node embeddings denoted as $\boldsymbol{E}'$. We then aggregate $\boldsymbol{E}$ and $\boldsymbol{E}'$ to enhance the representations of users and items in the target behavior,
\begin{equation}
  \label{eq:target_rep}
   \hat{\boldsymbol{E}} = \boldsymbol{E} \oplus \boldsymbol{E}',
\end{equation}
where $\oplus$ denotes the element-wise addition, $\hat{\boldsymbol{E}}$ represents users and items embeddings in the target behavior (matrix form), represented in embedding form as $\hat{\boldsymbol{e}}_u$ and $\hat{\boldsymbol{e}}_i$. When there is an insufficient record of user interaction in the target behavior, $\hat{\boldsymbol{e}}_u$ and $\hat{\boldsymbol{e}}_i$ contribute to enhancing the accuracy of recommendation. We will provide a detailed explanation in the next section.

\subsubsection{\textbf{Prediction \& Optimization}}
In this section, we will introduce the prediction and model optimization of the proposed model.

\textbf{\textit{Prediction}}. 
Currently, we have obtained two representations, $\tilde{\boldsymbol{e}}_{u}^{agg}$ and $\hat{\boldsymbol{e}}_u$, for user $u$ in the target behavior. $\tilde{\boldsymbol{e}}_{u}^{agg}$ represents the user's specific interests learned from individual interactions. It is characterized by a higher level of specificity and detail. $\hat{\boldsymbol{e}}_u$ represents similar preferences aggregated from high-order neighbors using GCN, and it tends to be more generalized in nature. When ample user interaction records exist in the target behavior, $\tilde{\boldsymbol{e}}_{u}^{agg}$ can more accurately represent user preferences. Conversely, in scenarios where there is a scarcity of user interaction records, the accuracy of $\tilde{\boldsymbol{e}}_{u}^{agg}$ cannot be guaranteed. In such cases, $\hat{\boldsymbol{e}}_u$ can provide enhanced recommendations by relying on preferences aggregated from higher-order neighbors. Hence, we configure hyperparameter $\lambda = 1/N_u$ to govern the influence of these two representations on recommendation, where $N_u$ is the number of user interactions in the target behavior. A comprehensive analysis is presented in Section~\ref{Parameter Analysis} of the experiment.

Following previous works~\cite{CRGCN, MBGCN, MB-CGCN}, we adopt the inner product score $y_{ui}$ to evaluate the user's interest in items to provide personalized recommendation,
\begin{equation}
  \label{eq:score}
   y_{ui} = (1- \lambda) \cdot (\tilde{\boldsymbol{e}}_{u}^{agg})^{\top} \boldsymbol{e}_i + \lambda \cdot (\hat{\boldsymbol{e}}_u)^{\top} \hat{\boldsymbol{e}}_i,
\end{equation}
where $\tilde{\boldsymbol{e}}_{u}^{agg}$ represents the user preference aggregated by the target behavior item-aware preferences of user $u$, $\boldsymbol{e}_i$ represents the embedding of item $i$ obtained through embedding pre-training. $\hat{\boldsymbol{e}}_u$ and $\hat{\boldsymbol{e}}_i$ correspond to the embeddings of user $u$ and item $i$ learned by the GCN enhancement module, respectively. 

\textbf{\textit{Optimization}}. Following the previous CF-based works~\cite{LightGCN, SMBREC, MB-CGCN}, the Bayesian Personalized Ranking (BPR) loss~\cite{BPR} is adopted for the recommendation performance optimization,
\begin{equation}
  \label{eq:bpr}
  \mathcal{L}_{bpr} = -\sum_{(u,i,j) \in \mathcal{O}} ln \sigma(y_{ui}-y_{uj}),
\end{equation}
where $\mathcal{O}=\{(u,i,j)|(u,i) \in \mathcal{R}^{+}, (u,j) \in \mathcal{R}^{-}\}$, and $\mathcal{R}^{+}$ and $\mathcal{R}^{-}$ represent the observed and unobserved interactions in the target behavior, respectively. 

Finally, we combine the above two objective functions and employ multi-task learning for joint optimization,
\begin{equation}
  \label{eq:total_loss}
  \mathcal{L} = \beta \cdot \mathcal{L}_{bce} + (1-\beta) \cdot \mathcal{L}_{bpr} + \gamma \cdot \left\lVert \boldsymbol \Theta \right\rVert_{2},
\end{equation}
where $\beta$ and $\gamma$ are hyperparameters, $\Theta$ represents all of the parameters in our model, and $\left\lVert \cdot \right\rVert_{2}$ denotes the $L_2$ regularization.
\section{experiment} \label{experiment}

\subsection{Experiment Settings}

\subsubsection{\textbf{Dataset}} Four real-world datasets that are used in our experiments:

\begin{itemize}[leftmargin=*]
  \item \textbf{Taobao.}\footnote{https://tianchi.aliyun.com/dataset/649} This dataset is an e-commerce website under the Alibaba Group. It encompasses three distinct user behaviors: \textit{click}, \textit{cart}, and \textit{purchase}, with the \textit{purchase} being the target behavior.

  \item \textbf{Tmall.}\footnote{https://tianchi.aliyun.com/dataset/140281} Similar to Taobao, Tmall is another e-commerce site owned by Alibaba. This dataset includes four types of user behaviors: \textit{click}, \textit{collect}, \textit{cart}, and \textit{purchase}, with the \textit{purchase} being the target behavior.
  
  \item \textbf{Yelp.}\footnote{https://www.kaggle.com/yelp-dataset/yelp-dataset} Yelp is a well-known review website for businesses in the United States. This dataset includes \textit{tips} and \textit{ratings}. Following previous works~\cite{KHGT, GNMR, MATN}, \textit{ratings} ($r$) are categorized into three behaviors: \textit{dislike} ($r \leq 2$), \textit{neutral} ($2<r<4$), and \textit{like} ($r \geq 4$), with \textit{like} being the target behavior. \textit{tips}, \textit{dislike}, and \textit{neutral} are considered as auxiliary behaviors.

  \item \textbf{ML10M.}\footnote{https://grouplens.org/datasets/movielens/10m/} MovieLens 10M dataset, which is a publicly available dataset developed by the GroupLens project for movie recommendation. Similar to the Yelp dataset, user ratings in this dataset are also categorized into three behaviors: \textit{dislike}, \textit{neutral}, and \textit{like}, with \textit{like} being the target behavior.

\end{itemize} 

For the four datasets, to eliminate duplicate data, we follow the previous works~\cite{NMTR, MBGCN} to retain only the earliest occurrence of each interaction. The specific statistical information of the four datasets is presented in Table~\ref{tab:dataset}.

\begin{table}[htb]
  \caption{Statistics of four datasets ("\# tar. avg." represents the average number of user interactions in the target behavior).}
  \label{tab:dataset}
  \resizebox{\columnwidth}{!}{
    \begin{tabular}{ccccccc}
      \toprule
      \textbf{Dataset} & \textbf{\# User} & \textbf{\# Item} & \textbf{\# Interation} & \textbf{\# tar. avg.}  & \textbf{Behavior Type} \\
      \midrule
      \textbf{Taobao} & 48,749 & 39,493 & $2.0\times10^6$   & 3    & \{\textit{click, cart, purchase}\} \\
      \textbf{Tmall}  & 41,738 & 11,953 & $2.3\times10^6$   & 4    & \{\textit{click, collect, cart, purchase}\} \\
      \textbf{Yelp}   & 19,800 & 22,734 & $1.4\times10^6$   & 33   & \{\textit{tips, dislike, neutral, like}\} \\
      \textbf{ML10M}  & 67,788 &  8,704 & $9.9\times10^6$   & 72   & \{\textit{dislike, neutral, like}\} \\
      \bottomrule
    \end{tabular}
  }
\end{table}

\subsubsection{\textbf{Evaluation Protocols \& Parameter Settings}}
To evaluate the effectiveness of our model, we employ two widely used evaluation metrics~\cite{MBGCN, MB-CGCN, PKEF}, \emph{Hit Ratio (HR@K)}~\cite{HR} and \emph{Normalized Discounted Cumulative Gain (NDCG@K)}~\cite{NDCG}, by considering the \textit{Top@K} recommended items. \emph{HR@K} focuses on the number of items that actually interacted with the user in the recommendation list, while \emph{NDCG@K} takes into account the ranking of items in the recommendation list. Besides, we utilize the leave-one-out validation method~\cite{NMTR, SMBREC, MBGCN} for model evaluation. 

For all methods, we uniformly set the batch size to 1024 and the embedding size to 64 during the training phase. Our model uses the Adam~\cite{Adam} optimizer for optimization. To determine the optimal values for the hyperparameters, including the learning rate, $\beta$, and $\gamma$, we perform a grid search on the sets $[1e^{-2}, 3e^{-3}, 1e^{-3}, 1e^{-4}]$, $[0.1, 0.3, 0.5, 0.7, 0.9]$ and $[1e^{-2}, 1e^{-3}, 1e^{-4}]$, respectively. For all GCN modules, we uniformly fixed the number of GCN layers at 2. For the method with cross-entropy loss function, we randomly select four negative samples for each positive sample. In addition, to ensure fairness, we also set parameters for the baselines according to the descriptions in their papers and perform a grid search to find the optimal values.

\subsubsection{\textbf{Baselines}}
To validate its effectiveness, we compare our model against nine baseline methods. Two of these are single-behavior methods, while the remaining seven are multi-behavior methods.

\textbf{Single-behavior models:}

\begin{itemize}[leftmargin=*]
  \item \textbf{MF-BPR}~\cite{BPR}. This is a matrix factorization-based recommendation approach. It computes user-item inner products for recommendations and utilizes the BPR loss function for optimization.
  \item \textbf{LightGCN}~\cite{LightGCN}. This method offers a simplified GCN-based approach, cutting out unnecessary non-linear designs. This simplification boosts recommendation performance while also reducing model complexity.
\end{itemize}

\begin{table*}[hbt]
  \caption{Overall performance across four datasets ("\textit{Impr.}" represents the relative performance improvement compared to the suboptimal metric, the bold font indicates the optimal metric, and the underline represents the suboptimal metric).}
  \label{tab:overall}
  \resizebox{\textwidth}{!}{
	\begin{tabular}{crccccccccccr}
	\toprule
	\multirow{2}{*}{\textbf{Dataset}} & \multirow{2}{*}{\textbf{Metric}} & \multicolumn{2}{c}{\textbf{Single-behavior}}       & \multicolumn{8}{c}{\textbf{Multi-behavior}}    & \multirow{2}{*}{\textit{\textbf{Impr.}}} \\ \cmidrule(lr){3-4} \cmidrule(lr){5-12}
	                      &                     & \textbf{MF-BPR} & \textbf{LightGCN} & \textbf{R-GCN} & \textbf{NMTR} & \textbf{MBGCN} & \textbf{S-MBRec} & \textbf{CRGCN} & \textbf{MB-CGCN} & \textbf{PKEF}  & \textbf{BCIPM}  & \\ \hline  
	\multirow{2}{*}{\textbf{Tmall}}   & \textbf{HR@10}   & 0.0230          & 0.0393       & 0.0316            & 0.0517         & 0.0549        & 0.0694         & 0.0840        & 0.1073     & \underline{0.1118}  & \textbf{0.1414} & 26.48\%  \\ 
					  & \textbf{NDCG@10} & 0.0124          & 0.0209       & 0.0157            & 0.0250         & 0.0285        & 0.0362         & 0.0442        & 0.0416     & \underline{0.0630}  & \textbf{0.0741} & 17.62\%  \\ \hline

	\multirow{2}{*}{\textbf{Taobao}}  & \textbf{HR@10}   & 0.0178          & 0.0254       & 0.0283            & 0.0409         & 0.0434        & 0.0571         & \underline{0.1152}        & 0.0989     & 0.1097  & \textbf{0.1292} & 12.15\%  \\ 
					  & \textbf{NDCG@10} & 0.0101          & 0.0138       & 0.0148            & 0.0212         & 0.0259        & 0.0331         & \underline{0.0629}        & 0.0470     & 0.0627  & \textbf{0.0716} & 13.83\%  \\ \hline

	\multirow{2}{*}{\textbf{Yelp}}    & \textbf{HR@10}   & 0.0327          & 0.0400       & 0.0348            & 0.0324         & 0.0356        & 0.0353         & 0.0367        & 0.0355     & \underline{0.0423}  & \textbf{0.0502} &  18.68\%  \\ 
					  & \textbf{NDCG@10} & 0.0159          & 0.0202       & 0.0176            & 0.0160         & 0.0183        & 0.0173         & 0.0178        & 0.0164     & \underline{0.0210}  & \textbf{0.0244} &  16.19\%  \\ \hline

	\multirow{2}{*}{\textbf{Ml10M}}   & \textbf{HR@10}   & 0.0585          & \underline{0.0662}       & 0.0552            & 0.0451         & 0.0469        & 0.0340         & 0.0502        & 0.0627     & 0.0591  & \textbf{0.0810} & 22.36\%  \\ 
					  & \textbf{NDCG@10} & 0.0274          & \underline{0.0319}       & 0.0262            & 0.0202         & 0.0228        & 0.0164         & 0.0239        & 0.0302     & 0.0264  & \textbf{0.0392} & 22.88\%  \\ 
	\bottomrule
	\end{tabular}
  }
\end{table*}

\textbf{Multi-behavior models:}

\begin{itemize}[leftmargin=*]
  \item \textbf{R-GCN}~\cite{R-GCN}. While R-GCN is not designed for recommendation tasks, its applicability to such domains is being demonstrated. The model has offered valuable insights for extracting information from heterogeneous graph networks.
  \item \textbf{NMTR}~\cite{NMTR}. NMTR is an approach that combines DNN with CF. It links different behaviors by sequentially propagating interaction scores and uses multi-task learning to optimize the model.
  \item \textbf{MBGCN}~\cite{MBGCN}. The model learns user preferences within behavior-specific graphs and combines them based on behavior importance. It also uses an item-item graph network to enhance recommendation capabilities by learning item embeddings.
  \item \textbf{S-MBRec}~\cite{SMBREC}. The model learns node representations within each behavior's graphs. It then combines user and item features separately and includes a star contrast learning module to explore similarities between different behaviors.
  \item \textbf{CRGCN}~\cite{CRGCN}. CRGCN presents a method for modeling behavior sequences. It uses a residual module based on GCN for each behavior and facilitates information transfer between modules through embedding propagation, establishing dependency relationships between behaviors.
  \item \textbf{MB-GCGN}~\cite{MB-CGCN}. This model simplifies CRGCN by using LightGCN directly to capture user preferences within each behavior and conducting feature transformation before information propagation.
  \item \textbf{PKEF}~\cite{PKEF}. This model builds on MB-CGCN by adding a parallel knowledge fusion module and a projection disentangling multi-expert network. These designs aim to address imbalanced data distribution in multiple behaviors.
\end{itemize}

\subsection{Overall Performance}

Table~\ref{tab:overall} showcases the comparative experimental results of our proposed model alongside nine benchmark baseline models, evaluated across four distinct datasets.

From the experimental results of the single-behavior methods, LightGCN consistently surpasses MF-BPR in performance across all four datasets. This superior performance is due to the enhanced representation learning capability inherent in Graph Convolutional Networks (GCNs). LightGCN is adept at assimilating information from high-order neighbors, in contrast to MF-BPR, which is constrained to deriving insights solely from direct interactions.

The multi-behavior models yielded better results than the single-behavior method on Tmall and Taobao, indicating the effectiveness of involving multiple behaviors. However, R-GCN underperforms LightGCN on both Yelp and Ml10M. This is because R-GCN incorporates user preferences from auxiliary behaviors by summing them directly without differentiating various behaviors. In NMTR, a sequential modeling approach establishes connections between behaviors by propagating prediction scores backward, leading to superior performance compared to the single-behavior method. CRGCN, also grounded in sequential modeling, facilitates more extensive collaborative information sharing via embeddings, thus outperforming NMTR. MBGCN exhibits a significantly superior performance compared to R-GCN by assigning distinct weights to different behaviors. 
S-MBREC has developed unique fusion strategies for both users and items, resulting in improved performance compared to MBGCN. Furthermore, both MBGCN and S-MBREC outperform R-GCN when auxiliary behaviors are included, indicating that noise is introduced and effective fusion strategies can help filter out such noise. Both MB-CGCN and PKEF are variations of CRGCN. MB-CGCN simplifies CRGCN while PKEF adds a parallel knowledge fusion module to MB-CGCN, which helps to address different issues. Both models have demonstrated outstanding performance on the two datasets, but each has specific strengths and weaknesses on different datasets.

After analyzing the experimental results on Yelp and ML10M, we observed an interesting trend: multi-behavior methods do not consistently improve performance and, in some cases, they perform worse than the single-behavior method, LightGCN. This phenomenon can be attributed to two key factors. First, the nature of multi-behavior data in Tmall and Taobao shows a strong correlation among different behaviors, which enables the extraction of more valuable information from auxiliary behaviors. In contrast, the Yelp and ML10M datasets exhibit a weaker correlation among behaviors, resulting in a diminished utility of information derived from auxiliary sources.
Second, there is a significant difference in the average number of user interactions within the target behaviors of different datasets. Specifically, Yelp and ML10M have higher interaction numbers, whereas Taobao and Tmall exhibit lower interaction numbers in the target behaviors (as detailed in Table~\ref{tab:dataset}). Consequently, integrating auxiliary behaviors can help mitigate data sparsity issues on Taobao and Tmall. However, in Yelp and ML10M scenarios, this integration tends to introduce more noise rather than valuable information. These observations align with our hypothesis that incorporating user preferences from auxiliary behaviors can lead to noise.

Our proposed BCIPM model sets a new benchmark in achieving state-of-the-art performance across all four datasets. These experimental results strongly substantiate its efficacy and versatility. The exceptional performance can be credited to its two-pronged design. The BIPN adeptly discerns user-specific and detailed preferences, while its GCN enhancement module adeptly understands users' general preferences. Furthermore, the incorporation of embedding pre-training is pivotal, ensuring a thorough learning of both detailed and general aspects of user preferences. 
To summarize, we have reached the following conclusions: (1) The incorporation of multiple behaviors substantially improves the performance of recommendation systems; (2) When integrating multiple behaviors, it is crucial to consider the potential issue of introducing noise; (3) Our model demonstrates robust adaptability and effectiveness across various datasets. In the forthcoming ablation study, we will delve into a detailed examination of the contribution and effectiveness of each individual module.

\subsection{Ablation Study}

\subsubsection{\textbf{Component Effectiveness Analysis}}
To evaluate the effectiveness of the individual design components in our proposed method, this section presents ablation experiments for each component. The specific experimental setups are as follows: (1) \textbf{Removing the embedding pre-training module} (\textit{\textbf{w/o. pre.}}): In this experiment, the embedding pre-training module is excluded, and randomly initialized embeddings are directly used as the input of the downstream modules. (2) \textbf{Removing the GCN enhancement module} (\textit{\textbf{w/o. enh.}}): In this experiment, we exclude the GCN enhancement module and only aggregate the learned item-aware preferences from target behavior for recommendation. (3) \textbf{Removing the behavior-contextualized item preference network module} (\textit{\textbf{w/o. net.}}): In this experiment, we remove the BIPN module and only retain the user and item embeddings learned by the GCN enhancement module for recommendation. The experimental results are reported in Table~\ref{tab:abl}.

\begin{table}[t]
  \caption{Ablation study of the individual design components ("\textit{w/o. pre.}", "\textit{w/o. inf.}" and "\textit{w/o. net.}" represent methods that without embedding pre-training module, without GCN enhancement module, and without BIPN module, respectively).}
  \label{tab:abl}
  \resizebox{\columnwidth}{!}{
    \begin{tabular}{crcccc}
    \toprule
    \textbf{Dataset} & \textbf{Metric} & \textit{\textbf{w/o. pre.}} & \textit{\textbf{w/o. enh.}} & \textit{\textbf{w/o. net.}} & \textit{\textbf{our}}  \\ \cmidrule(lr){1-2} \cmidrule(lr){3-5} \cmidrule(lr){6-6}
    \multirow{2}{*}{\textbf{Tmall}}     & \textbf{HR@10}   & 0.0559   & 0.0863                      & 0.1043                      & \textbf{0.1414} \\
					& \textbf{NDCG@10} & 0.0277   & 0.0405                      & 0.0580                      & \textbf{0.0741} \\ \hline
    \multirow{2}{*}{\textbf{Taobao}}    & \textbf{HR@10}   & 0.0346   & 0.1025                      & 0.1147                      & \textbf{0.1292} \\
					& \textbf{NDCG@10} & 0.0179   & 0.0560                      & 0.0667                      & \textbf{0.0716} \\ \hline
    \multirow{2}{*}{\textbf{Yelp}}      & \textbf{HR@10}   & 0.0400   & 0.0343                      & 0.0296                      & \textbf{0.0502} \\
                                        & \textbf{NDCG@10} & 0.0195   & 0.0159                      & 0.0132                      & \textbf{0.0244} \\ \hline
    \multirow{2}{*}{\textbf{ML10M}}     & \textbf{HR@10}   & 0.0775   & 0.0803                      & 0.0710                      & \textbf{0.0810} \\
                                        & \textbf{NDCG@10} & 0.0372   & 0.0374                      & 0.0327                      & \textbf{0.0392} \\
    \bottomrule
    \end{tabular}
  }
  \vspace{-10pt}
\end{table}

For the method \textit{\textbf{w/o. pre.}}, there is a remarkable decrease in performance on the Tmall and Taobao datasets. Our analysis suggests that this decline is associated with the number of user interactions in the datasets. Specifically, in the Tmall and Taobao datasets, where there are few user interaction records in the target behavior (see Table~\ref{tab:dataset}), leading to challenges in mining sufficient useful information. This result further confirms the importance of embedding pre-training, particularly in situations with limited interaction data. It enables the utilization of auxiliary behaviors to gain more collaborative information.
When there are sufficient user interaction records in the target behavior, the behavior-contextualized item preference network can learn user-specific and detailed preferences, leading to more accurate recommendations. Therefore, in the Yelp and ML10M datasets, \textit{\textbf{w/o. enh.}} performs better than \textit{\textbf{w/o. net.}}. However, when there are fewer user interaction records in the target behavior, the learning capability of the behavior-contextualized item preference network may be insufficient. In the Tmall and Taobao datasets, the GCN enhancement module can improve recommendation quality by aggregating information from high-order neighbors. This demonstrates the superior performance of \textit{\textbf{w/o. net.}}. Our approach (see "\textit{\textbf{our}}") surpasses both \textit{\textbf{w/o. enh.}} and \textit{\textbf{w/o. net.}} in all four datasets, benefiting from the combination of the two types of preferences mentioned above. A detailed discussion will be provided in Section~\ref{Parameter Analysis}.

In summary, the experimental results indicate that the removal of any individual module from the model leads to a decline in performance. These results demonstrate the effectiveness and essential contribution of each component within the model's design.

\subsubsection{\textbf{Behavior-Contextualized Item Preference Network Analysis}}
\begin{figure}[t]
    \centering
    \includegraphics[width=1.\linewidth]{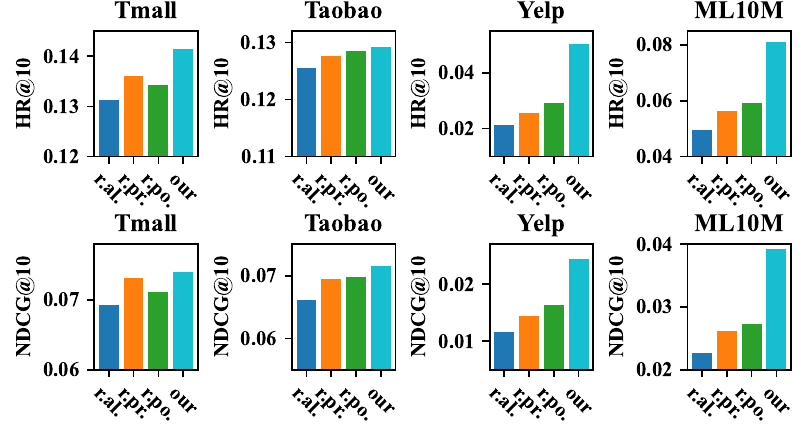}
    \caption{Behavior-contextualized item preference network analysis ("\textbf{\textit{r.al.}}" indicates remove both pre-filtering layer and post-filtering layer, "\textbf{\textit{r.pr.}}" means remove pre-filtering layer, "\textbf{\textit{r.po.}}" means remove post-filtering layer, and "\textbf{\textit{our}}" is our method).}
    \label{fig:network}
\end{figure}
To verify the effectiveness of the BIPN module, we performed ablation experiments on its components. Specifically: (1) Simultaneously removing the pre-filtering layer and the post-filtering layer (denoted as \textbf{\textit{r.al.}}). (2) Solely removing the pre-filtering layer (denoted as \textbf{\textit{r.pr.}}). (3) Solely removing the post-filtering layer (denoted as \textbf{\textit{r.po.}}). The results of these ablation experiments are illustrated in Figure~\ref{fig:network}.

The experimental results consistently indicate that in all four datasets, whether removing the pre- or post-filtering layer alone (see \textbf{\textit{r.pr.}} and \textbf{\textit{r.po.}}) or both simultaneously (see \textbf{\textit{r.al.}}), the model's performance significantly decreased. This is because the filtering layers can effectively filter noise based on the embedding triplet <\textit{user, item, behavior}>. Compared to the results of removing a single layer of filtering (see \textbf{\textit{r.pr.}} and \textbf{\textit{r.po.}}), the results of simultaneously removing both pre- and post-filtering layers (see \textbf{\textit{r.al.}}) are the worst, indicating that both pre- and post-filtering layers are necessary. In comparison to our method (see \textbf{\textit{our}}), the performance declines after removing the filtering layers (see \textbf{\textit{r.pr.}}, \textbf{\textit{r.po.}} and \textbf{\textit{r.al.}}) in the Tmall and Taobao datasets are less severe than in the Yelp and ML10M datasets. It can be observed that the Yelp and ML10M datasets are more sensitive to changes in the network components due to the allocation of more weight to the BIPN module. This allocation is determined by the quantity of user interactions in the target behavior. On the other hand, the Tmall and Taobao datasets are less susceptible to internal variations as they allocate less weight to this component.

\subsubsection{\textbf{Effectiveness Analysis of Auxiliary Behavior}}

\begin{figure}[t]
    \centering
    \includegraphics[width=1.\linewidth]{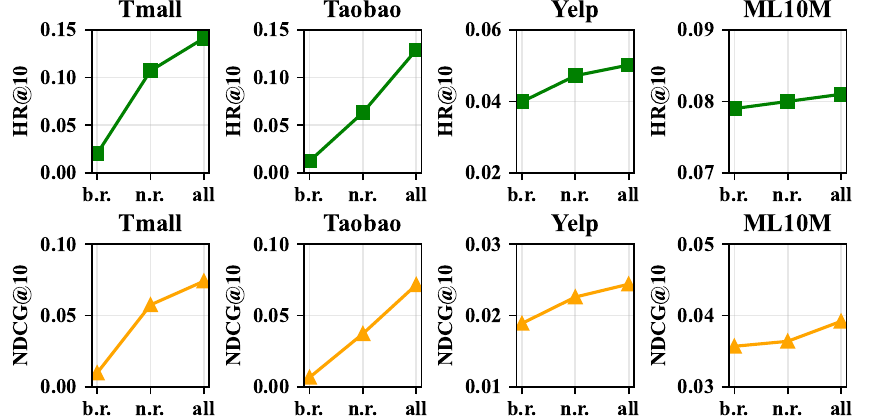}
    \caption{Effectiveness analysis of auxiliary behavior ("\textbf{\textit{b.r.}}" indicates simultaneously eliminate auxiliary behavior from both two modules, "\textbf{\textit{n.r.}}" means exclude auxiliary behaviors solely from the BIPN module, and "\textbf{\textit{all}}" implies keep them all).}
    \label{fig:behaviors}
\end{figure}
In our approach, auxiliary behavior is introduced in both the embedding pre-training module and the BIPN module. To verify the effectiveness of incorporating auxiliary behavior, we conducted two experiments: (1) Simultaneously eliminate auxiliary behavior from both the embedding pre-training module and the BIPN module (denoted as \textbf{\textit{b.r.}}). (2) Exclude auxiliary behaviors solely from the BIPN module while retaining them in the embedding pre-training module (denoted as \textbf{\textit{n.r.}}). The experimental results are presented in Figure~\ref{fig:behaviors}.
According to the experimental results, the exclusion of auxiliary behaviors from both the embedding pre-training module and the BIPN module leads to a substantial decrease in performance across all four datasets, thus confirming the necessity of modeling auxiliary behavior data. 

Our experiments on Tmall and Taobao showed that removing auxiliary behaviors from both modules simultaneously led to a drastic reduction in the model's predictive capability. This is largely attributed to the sparsity issue of target behavior data within these datasets. The integration of auxiliary behavior data during the embedding pre-training phase allows for the extraction of more collaborative information from an expanded graph. Similarly, incorporating auxiliary behaviors into the BIPN module broadens the training dataset and enhances the optimization of network parameters. In stark contrast, the Yelp and ML10M datasets, which are characterized by a richer set of interaction data in the target behavior, exhibit a more gradual decline in performance following the removal of auxiliary behaviors.

\subsubsection{\textbf{Embedding Pre-training Analysis}} \label{Strategy Analysis}

In Section~\ref{Embedding Pre-training}, we introduced two distinct methods for embedding pre-training. This section is dedicated to presenting an experimental comparison between these methods: (1) Performing GCN operations separately in different behaviors and then summing the embeddings obtained from each behavior as the pre-trained embedding (denoted as \textit{\textbf{sep.}}). (2) Employing an aggregated graph constructed from all behavior data without distinguishing behavior types (denoted as \textit{\textbf{agg.}}). The experimental results are reported in Table~\ref{tab:pre-traing}.

\begin{table}[t]
  \caption{Embedding pre-training strategy components ("\textit{\textbf{sep.}}" represents performing GCN individually on each behavior and subsequently aggregating the results, "\textit{\textbf{agg.}}" means performing GCN on the graph constructed from all behavior data without distinguishing behavior types).}
  \label{tab:pre-traing}
  \resizebox{0.6\columnwidth}{!}{
    \begin{tabular}{crcc}
    \toprule
    \textbf{Dataset} & \textbf{Metric} & \textit{\textbf{sep.}} & \textit{\textbf{agg.}}   \\  \cmidrule(lr){1-2} \cmidrule(lr){3-3} \cmidrule(lr){4-4}
    \multirow{2}{*}{\textbf{Tmall}}     & \textbf{HR@10}   & 0.0948   & \textbf{0.1414} \\
					                              & \textbf{NDCG@10} & 0.0505   & \textbf{0.0741} \\ \hline
    \multirow{2}{*}{\textbf{Taobao}}    & \textbf{HR@10}   & 0.0919   & \textbf{0.1292} \\
					                              & \textbf{NDCG@10} & 0.0548   & \textbf{0.0716} \\ \hline
    \multirow{2}{*}{\textbf{Yelp}}      & \textbf{HR@10}   & 0.0376   & \textbf{0.0502} \\
                                        & \textbf{NDCG@10} & 0.0164   & \textbf{0.0244} \\ \hline
    \multirow{2}{*}{\textbf{ML10M}}     & \textbf{HR@10}   & 0.0722   & \textbf{0.0810} \\
                                        & \textbf{NDCG@10} & 0.0332   & \textbf{0.0392} \\
    \bottomrule
    \end{tabular}
  }
  \vspace{-10pt}
\end{table}

The results show that our adopted method~\textit{\textbf{agg.}} is significantly superior to the method of conducting GCN operations separately for each behavior and subsequently aggregating (\textit{\textbf{sep.}}).  The marked superiority of \textit{\textbf{agg.}} can be attributed to its alignment with the core objective of pre-training, which is to harness a broader scope of collaborative information without being constrained by specific types of behaviors. In \textit{\textbf{sep.}}, there is a potential for information loss during the aggregation process due to the variations in user preferences across different behaviors. Conversely, our approach, which amalgamates data without distinguishing between behavior types, utilizes a graph that integrates interaction data from all behaviors. This integration results in a graph with more interconnected nodes, facilitating the extraction of a more comprehensive array of collaborative information.

\subsection{Parameter Analysis} \label{Parameter Analysis}

\begin{figure}[t]
    \centering
    \includegraphics[width=1.\linewidth]{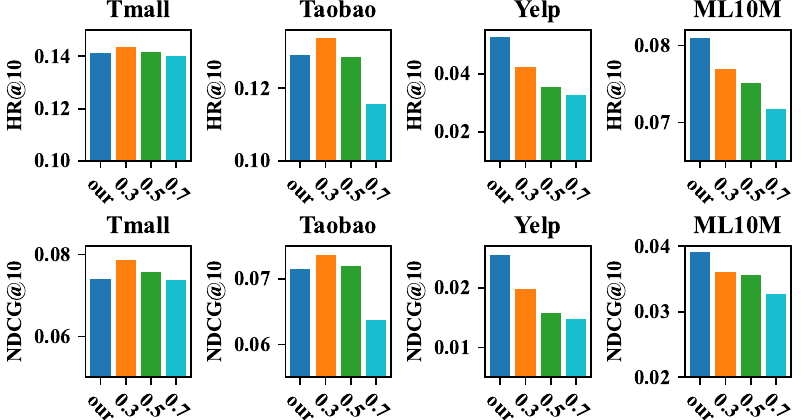}
    \caption{Influence analysis of $\lambda$ ("\textit{our}" indicates the configuration parameter used in BCIPM, and the values "0.3", "0.5", and "0.7" represent the corresponding outcomes when $\lambda$ is set to 0.3, 0.5, and 0.7, respectively).}
    \label{fig:lambda}
    \vspace{-15pt}
\end{figure}

To assess the impact of the hyperparameter $\lambda$ on the performance of our model, we conducted a series of experiments with $\lambda$ set at values of 0.3, 0.5, and 0.7. These experiments were aimed at evaluating the performance variations of the BCIPM under different $\lambda$ settings. The results of these experiments are depicted and analyzed in Figure~\ref{fig:lambda}.

In the BCIPM, assigning a larger value to the hyperparameter $\lambda$ implies an increased emphasis on the role of the GCN enhancement module. In contrast, a smaller $\lambda$ value highlights the significance of the BIPN module. Our experiments across all four datasets reveal a consistent trend: as $\lambda$ increases, there is a noticeable decline in BCIPM's performance, particularly in terms of HR@10 and NDCG@10 metrics. This trend underscores the BIPN module as the primary driver of performance improvement. This observation is in perfect harmony with the underlying design of our proposed approach, which prioritizes the capture of item-aware preferences in the target behavior through the BIPN module, relegating the GCN enhancement module to a more ancillary role.

It is essential to recognize that while the selection of the hyperparameter $\lambda$ in BCIPM yielded superior performance on Yelp and ML10M, this was not the case for Tmall and Taobao. This variation in outcomes can be ascribed to the differences in user-item interaction frequencies across these datasets. Notably, the Tmall and Taobao datasets exhibit a relatively low average number of interactions between users and items (4 and 3, respectively, as detailed in Table~\ref{tab:dataset}). In our approach, we approximately set $\lambda$ as the inverse of the number of items with which users interact in the target behavior.
This heuristic, however, resulted in a sub-optimal allocation of the hyperparameter for numerous users, particularly in contexts where item-aware preferences are insufficient to drive effective recommendations due to limited user-item interactions. In such scenarios, it becomes imperative to assign greater significance to the preferences derived from the GCN enhancement module, ensuring a more robust and effective recommendation.


\section{conclusion} \label{conclusion}

 In this paper, we revisit the utilization of multi-behavior data in recommendation methods and introduce a novel approach for multi-behavior recommendation. Specifically, our approach employs a two-pronged design, with the behavior-contextualized item preference network designed to learn item-aware preferences and the GCN enhancement module capturing high-order neighbor preferences of users. The combination of these two distinct preferences leads to a significant improvement in recommendation performance. In addition, an embedding pre-training module is adopted to facilitate embedding learning. This module can help users learn item-aware preferences and capture high-order neighbor preferences effectively. Extensive experiments are conducted on four real-world datasets and the experiment results demonstrate the effectiveness of our approach. Moving forward, our research will focus on devising an advanced weight allocation strategy for the GCN enhancement module, to further improve the performance of recommendation model.

\begin{acks}
  This research is supported by the National Natural Science Foundation of China under Grant 62376186, 61932009 and 62272254.
\end{acks}

\bibliographystyle{ACM-Reference-Format}
\balance
\bibliography{reference}

\end{document}